\documentclass[12pt]{article}
\usepackage{amssymb}
\usepackage{amsmath}
\usepackage{hyperref}
\begin{document}
\title{\bf   Holographic Aspects of Non-minimal $RF^{(a)}_{\mu \alpha }F^{(a)\mu \alpha} $ Black Brane}
\author{ 
      Mehdi Sadeghi\thanks{Email:  mehdi.sadeghi@abru.ac.ir}\hspace{2mm}\\
{\small {\em Department of Physics, School of Sciences,}}\\
        {\small {\em Ayatollah Boroujerdi University, Boroujerd, Iran}}
       }
\date{\today}
\maketitle

\abstract{  In this paper, we consider Einstein-Hilbert gravity in the presence of cosmological constant and an electric field of Yang-Mills type, which is minimally coupled to gravity. We couple the Ricci scalar to the Yang-Mills invariant to obtain a modified theory of gravity. The black brane solution of this model is introduced up to the first order of the $RF^{(a)}_{\mu \alpha }F^{(a)\mu \alpha} $ term. Then, the color non-abelian direct current (DC) conductivity and the ratio of shear viscosity to entropy density are calculated for this solution. Our results recover the Yang-Mills Schwarzschild AdS black brane in the limit of $q_2 \to 0$.}\\

\noindent PACS numbers: 11.10.Jj, 11.10.Wx, 11.15.Pg, 11.25.Tq\\

\noindent \textbf{Keywords:}    AdS/CFT duality, DC Conductivity, Black brane , Shear viscosity to entropy density 

\section{Introduction} \label{intro}
Black holes are solutions to the Einstein equations that possess event horizon(s). An event horizon is a null hypersurface where there is no causal relation between the two sides of it. The fact that nothing can escape from the event horizon is the classical description of a black hole. Therefore, the key question is how to study the interior of a black hole. The radiation of a black hole was discovered by Hawking and Page [\cite{Hawking1983}]. This discovery, known as the quantum description of black holes, allows us to study the interior of these celestial objects. So, the quantum aspects of black holes are a key question that we are interested in studying.\\

The cosmological constant, which describes the expansion of the Universe, is positive in cosmology. The solution of Einstein's equations with a positive cosmological constant is known as de Sitter (dS) spacetime. The theory of gravity with a negative cosmological constant is known as Anti-de Sitter (AdS) spacetime, and it has a dual theory on the boundary. This proposal, introduced by Maldacena, is called AdS/CFT duality [\cite{Maldacena}]-[\cite{ Aharony}]. This is a powerful tool for describing strongly coupled field theories. This duality in the low energy limit is called fluid-gravity duality \cite{Son}-\cite{Bhattacharyya}. The theory on the boundary is described by hydrodynamics. Hydrodynamics is an effective description of field theory in the low-energy limit. Determining the transport coefficients helps us study fluid theory more effectively. In this paper, we utilize the Green-Kubo formula \cite{Son} to calculate the transport coefficients. The color non-abelian DC conductivity and shear viscosity to entropy density are investigated in this paper.\\

The motivation behind non-minimal theory, that is defined by coupling the gravitational field to other fields, is to propose an alternative theory of gravity. The non-minimal theory has five classes, and in this paper, we are interested in coupling the Ricci scalar to the Yang-Mills invariant [\cite{Balakin:2015gpq},\cite{Edery:2018jyp}]. Modified Maxwell-$F(R)$ gravity, as a non-minimal theory, can explain inflation and the late-time acceleration of the Universe\cite{Bamba:2008ja}. Power-law inflation can be understood through the non-minimal gravitational coupling of the electromagnetic field. Therefore, dark matter and dark energy, which are unknown components of the universe, can be explained by a non-minimal model. These non-minimal theories construct exact solutions of stars \cite{Horndeski:1978ca}, \cite{Mueller-Hoissen:1988cpx}, wormholes \cite{Balakin:2007xq}, \cite{Balakin:2010ar}, BHs \cite{Balakin:2007am}, \cite{Balakin:2015oea} and regular magnetic BHs \cite{Balakin:2006gv} with Wu-Yang ansatz. Therefore, studying the holographic aspects of this model can reveal some physical information.\\
Our model is non-abelian since the strong and weak interaction are non-abelian. There is some evidence that quark-gluon-plasma (QGP) existed at the beginning of the Universe. Therefore, it motivates us to study non-abelian term in presence of Einstein-Hilbert gravity to reveal some information of Big Bang.\\

Conductivity and shear viscosity to entropy density are bounded by a universal value.\\

Kovtun, Son and Starinets (KSS) bound states that $\frac{\eta}{s} \geq \frac{1}{4 \pi} $ for all quantum field theories. This bound is saturated for Einstein-Hilbert gravity with a field theory dual and it is violated for higher derivative gravities\cite{Brigante:2007nu}, massive gravity theory \cite{Sadeghi:2015vaa}-\cite{Parvizi:2017boc}, hairy anti-de-Sitter black hole solutions in generalized scalar-tensor gravity\cite{Bravo-Gaete:2020lzs}, planar hairy black hole configurations for a special subclass of the Horndeski theory\cite{Bravo-Gaete:2021hlc}, degenerate-higher-order-scalar-tensor theories\cite{Bravo-Gaete:2022lno}. We mention that this value is in good agreement with the experimental data of quark-gluon plasma (QGP). This evidence supports the reliability of string theory.\\

The bound of DC conductivity is given by $\sigma \geq \frac{1}{e^2}=1$ where $e$ represents the unit of charge carried by the gauge field- not the unit of charge in the boundary of theory. This bound is violated in the context of massive gravity \cite{Grozdanov:2015qia}, theories with background fields \cite{Donos:2014cya}, Non-abelian Einstein-Born-Infeld AdS theory\cite{Sadeghi:2021qou}, and AdS black brane coupled to non-abelian logarithmic gauge theory\cite{Sadeghi:2022mog}.\\

In this paper, we aim to study non-minimal black branes with a cosmological constant and investigate their holographic aspects through the calculation of conductivity and shear viscosity to entropy density. Then, we want to investigate whether these bounds are preserved for our model or not.
\section{    Non-minimal $RF^{(a)}_{\mu \alpha }F^{(a)\mu \alpha} $  AdS Black Brane Solution}
\label{sec2}
\indent The non-minimal Einstein-Yang-Mills theory with negative cosmological constant can be described in terms of the action functional below\cite{Balakin:2015gpq},\cite{Sert:2020vmq},\cite{Lambiase:2008zz},
\begin{eqnarray}\label{action}
S=\int d^{4}  x\sqrt{-g} \bigg[\frac{1}{\kappa }(R-2\Lambda )+\frac{q_1}{2}F^{(a)}_{\mu \alpha }F^{)(a)\mu \alpha} +q_2 RF^{(a)}_{\mu \alpha }F^{(a)\mu \alpha} \bigg],
\end{eqnarray}
where $\kappa$ is the gravitational constant, $R$ is the Ricci scalar, $\Lambda=-\frac{3}{L^2}$ is the cosmological constant, $L$ is the AdS radius,  $q_1$ is the constant parameter and dimensionless, $\mathcal{F}={\bf{Tr}}( F_{\mu \nu }^{(a)} F^{(a)\, \, \mu \nu })$  is Yang-Mills invariant, and index "$a$" refers to gauge group.\\
$ F^{(a)\, \, \mu \nu }$ is the Yang-Mills field tensor,
\begin{align} \label{YM}
F^{(a)}_{\mu \nu } =\partial _{\mu } A^{(a)}_{\nu } -\partial _{\nu } A^{(a)}_{\mu } -if^{(a)}_{(b)(c)  }[A^{(a)}_{\mu }, A^{(b)}_{\nu }],
\end{align}
in which the gauge coupling constant ,$f^{(a)}_{(b)(c) } $, is set to  1, and $A_{\nu }$'s represents the Yang-Mills potentials. $q_2$ is the dimensionful coupling constant and it is also interaction term between gauge field and the Ricci scalar.\\
Variation of the action (\ref{action}) with respect to the spacetime metric $g_{\mu \nu } $  yields the field equations,
\begin{equation}\label{EOM1}
R_{\mu \nu }-  \tfrac{1}{2} g_{\mu \nu } R + \Lambda g_{\mu \nu }=\kappa T^{\text{(eff)}}_{\mu \nu }
\end{equation}
where,
\begin{equation}
T^{\text{(eff)}}_{\mu \nu }=q_1T^{\text{(YM)}}_{\mu \nu } + q_2T^{(I)}_{\mu \nu }
\end{equation}
\begin{equation}
T^{\text{(YM)}}_{\mu \nu }=  \tfrac{1}{4}g_{\mu \nu } F^{(a)}_{\alpha \beta } F^{^{(a)} \alpha \beta }-F_{\mu }{}^{(a)\alpha } F^{(a)}_{\nu \alpha }
\end{equation}
\begin{eqnarray}
&&T^{(I)}_{\mu \nu }=\tfrac{1}{2} F^{(a)}_{\alpha \beta } F^{(a)\, \alpha \beta} g_{\mu \nu  } R-R_{\mu \nu } F^{(a)}_{\alpha \beta } F^{^{(a)} \alpha \beta } - 2 F_{\mu }^{(a) \,\alpha} F^{(a)}_{\nu \alpha }  R \nonumber \\ 
&&-2 F^{(a)}_{\alpha \beta }  g_{\mu \nu  }\nabla_{\gamma }\nabla^{\gamma }F^{(a)}_{\alpha \beta }-2 g_{\mu \nu  } \nabla_{\gamma }F^{(a)} _{\alpha \beta } \nabla^{\gamma }F^{(a) \,\alpha \beta } + F^{(a) \,\alpha \beta } \nabla_{\mu } \nabla_{\nu } F^{(a)}_{\alpha \beta } \nonumber \\ 
&& +2 \nabla_{\mu } F^{(a) \,\alpha \beta } \nabla_{\nu } F^{(a) \,\alpha \beta }+F^{(a) \,\alpha \beta } \nabla_{\nu } \nabla_{\mu } F^{(a)}_{\alpha \beta }.
\end{eqnarray}
Variation of the action (\ref{action}) with respect to the  $A_{\mu}$ yields the field equations,
 \begin{eqnarray}\label{EOM-Maxwell}
\nabla_{\mu }\Big(q_1 F^{(a)\mu \nu } + 2 q_2 F^{(a)\mu \nu } R\Big)=0.
\end{eqnarray}
Since the space-time of our model is 4-dimensional with a two-dimensional spatial metric or line element, we consider the  following form as an ansatz for the metric,\\
\begin{equation}\label{metric}
ds^{2} =-e^{-2H(r)}f(r)dt^{2} +\frac{dr^{2} }{f(r)} +\frac{r^2}{L^2}(dx^2+dy^2).
\end{equation}
We apply the ansatz \ref{metric} for solving Eq.(\ref{EOM-Maxwell}) where the potential 1-form is expressed by,
\begin{equation}\label{background}
{\bf{A}}^{(a)} =\frac{i}{2}h(r)dt\begin{pmatrix}1 & 0 \\ 0 & -1\end{pmatrix},
\end{equation}
the gauge group is the diagonal generator of the Cartan subalgebra of $SU(2)$ \cite{Shepherd:2015dse}. The Cartan-Killing metric is $Tr(K_a K_b)=g_{ab}$. Where $K_i$ denote the generators in the adjoint representation.\\
Now, we can write the field equations of motion by consideration of Eqs.(\ref{EOM-Maxwell}-\ref{background}).\\
The $tt$-component of the field equations of motion is as follows,\\
\begin{eqnarray}\label{tt-comp}
 && r h' \Big(h'\left(4 q_2 f' \left(4 r H'-3\right)+r \left(q_1-4 q_2 
f''\right)\right)+4 q_2 r f' h''\Big)  \nonumber\\&&-\frac{2 e^{-2 H} \left(r \left(f'+\Lambda 
	r\right)+f\right)}{\kappa}+ 4 q_2 f \Bigg(h'^2 \left(2 r \left(H' \left(r H'+4\right)+2 r
H''\right)-1\right) \nonumber\\&&+2 r h' \left(h'' \left(4 r H'+2\right)+r h^{(3)}(r)\right)+2 r^2 h''^2\Bigg)=0,
\end{eqnarray}
for $q_2=0$ we have,
\begin{equation}\label{ttq20}
2 r e^{-2H} \left(f'+\Lambda  r\right)-q_1 \kappa r^2  h'^2+2 f e^{-2H}=0,
\end{equation}
The $rr$-component of the field equations of motion is as follows,\\
\begin{eqnarray}\label{rr-comp}
&&\kappa e^{2 H} h' \Biggl(r \bigg(h' \left(4 q_2 \left(f' \left(3-4 r H'\right)+r f''\right)-q_1 r\right)-4
q_2 r f' h''\bigg) \nonumber\\&&+4 q_2 f \Bigg(2 r h'' \left(r H'-2\right)+h' \left(2 r \left(H'\left(2 r
H'-3\right)-r H''\right)+1\right)\Bigg)\Biggl)\nonumber\\&&+2 r \left(f'+\Lambda  r\right)+2 f \left(1-2 r H'\right)=0,
\end{eqnarray}
in which $rr$- part of Einstein equations for $q_2=0$ is the same as Eq.(\ref{ttq20}).\\
Non-zero component of Eq.(\ref{EOM-Maxwell}) is as follows,\\
\begin{equation}
B_1(r) h''(r)+B_2(r)h'(r)=0
\end{equation}
where,
\begin{equation}
B_1(r)=r \left(2 q_2 f' \left(3 r H'-4\right)+r \left(q_1-2 q_2 f''\right)\right)+4 q_2 f \left(r^2 H''-\left(r
H'-1\right)^2\right),
\end{equation}
\begin{eqnarray}
&& B_2(r)=r \left(r H' \left(4 q_2 f''+q_1\right)+2 \left(-6 q_2 f''-q_2 r f^{(3)}(r)+q_1\right)\right)\nonumber\\&&+2
q_2 f' \left(r \left(H' \left(r H'+6\right)+5 r H''\right)-6\right)\nonumber\\&&+4 q_2 f(r) \left(H' \left(1-r^2
H''\right)-r^2 H'^3+r \left(4 H''+r H^{(3)}(r)\right)\right).
\end{eqnarray}
The solution of $h(r)$ is as,
\begin{equation}
h(r)= C_1\int^{r}\frac{e^{-H(u)}}{B(u)}du+C_2,
\end{equation}
\begin{eqnarray}
&&B(u)=6 q_2 u^2 f' H'-2 q_2 u^2 f''-8 q_2 u f'-4 q_2 u^2 f H'^2\nonumber \\&&+4 q_2 u^2 f H''+8 q_2 u f H'-4 q_2 f+q_1 u^2.
\end{eqnarray}
These equations do not have analytical solutions. Therefore, we will solve the metric and the gauge field equations perturbatively up to first order in $q_2$ as \cite{Myers:2009ij}-\cite{Dey:2015ytd}.\\
We consider the following forms for $f (r)$, $H(r)$ and $h (r)$.
\begin{equation}\label{f}
f(r)=f_0(r)+q_2 f_1(r),
\end{equation}
\begin{equation}\label{h}
h(r)=h_0(r)+q_2 h_1(r),
\end{equation}
\begin{equation}\label{H}
H(r)=H_0(r)+q_2 H_1(r),
\end{equation}
where $f_0(r)$, $h_0(r)$ and $H_0(r)$ are the leading order solutions of Einstein-Yang-Mills AdS black brane in four dimensions.\\
The $h_0(r)$ , $f_0(r)$ and $H_0(r)$ are found exactly as,
\begin{equation}
h_0(r)=C_2-C_1\int^r\frac{ 1}{q_1 u^2}du=Q(\frac{1}{r}-\frac{1}{r_h}),
\end{equation}
where $Q$ is the Yang-Mills charge,  $C_1=q_1 Q$ and $C_1=-C_2$,
\begin{equation}\label{f0}
f_0(r)=\frac{2M}{r}-\frac{\Lambda r^2}{3}+\frac{\kappa q_1}{2 r}\int^r  u^2 
h_0'^2 du=\frac{2M}{r}+\frac{r^2}{L^2}-\frac{\kappa q_1 Q^2}{2 r^2},
\end{equation}
\begin{equation}
H_0(r)=0.
\end{equation}
Blackenig factor on the event horizon should  be suppressed,$f(r_h)=0$ .  $M$ is mass of black brane and it can be fixed by applying this condition,\\
\begin{equation}\label{m}
M=\frac{\kappa q_1 Q^2}{4 r_h}-\frac{r_h^3}{2L^2}.
\end{equation}
By plugging Eq.(\ref{m})  in Eq.(\ref{f0}) we have,
\begin{equation}
f_0(r)=\frac{r^2}{L^2}(1-\frac{r_h^3}{r^3})-\frac{\kappa q_1 Q^2}{2 r}(\frac{1}{r}-\frac{1}{r_h}).
\end{equation}
Eq.(\ref{tt-comp}) and Eq.(\ref{rr-comp}) should be the same up to first order of $q_2$. Therefore, $H_1$ is calculated as follows,
\begin{equation}
H_1(r)=C_3+ 2  \kappa r h_0' h_0''- \kappa h_0'^2=-\frac{5 \kappa Q^2}{r^4}+C_3,
\end{equation}
where $C_3$  is dimensionless integration constant. Since the metric of our model is flat near the boundary $r \to \infty$ [\cite{Mahapatra:2016dae}], so $C_3=0$.
It also guarantees the speed of light to unity on the theory of boundary.\\
\begin{equation}
h_1(r)=C_4+C_5\int^r \frac{ 8 u f_0'+2 u^2 f_0''+4 f_0-q_1 u^2 H_1}{q_1^2 u^4}du=C_4-\frac{ \kappa C_5 Q^3 }{ q_1 r^5}-\frac{24 C_5 Q }{L^2 q_1^2 r}
\end{equation}
where $C_5=q_1^2 Q$. $C_4$ is determined by applying this condition $h_1(r_h)=0$.
\begin{equation}
h_1(r)=-\kappa q_1  Q^3(\frac{1}{r^5}-\frac{1}{r_h^5})-\frac{24 Q }{L^2  }(\frac{1}{r_h}-\frac{1}{r}),
\end{equation}
by substituting $H_0(r)=0$ we have,
\begin{equation}
f_1(r)=\frac{1}{r}\int ^r D(u)du+\frac{C_6}{r},
\end{equation}
where,
\begin{eqnarray}
&&D(u)= -2 \kappa u^2 f_0'' h_0'^2+2 \kappa u^2 f_0' h_0' h_0''-6 \kappa u f_0'
h_0'^2+8 \kappa u f_0 h_0' h_0''\nonumber \\&&-2 \kappa f_0 h_0'^2+2 u f_0
H_1'+\kappa q_1 u^2 h_0' h_1'+\kappa q_1 u^2 H_1 h_0'^2,
\end{eqnarray}
\begin{equation}\label{f1}
f_1(r)=\frac{C_6}{r}+\frac{2\kappa Q
	^2}{L^2  r^2}-\frac{24 \kappa Q^2}{L^2  r^3}-\frac{7 \kappa   Q^2 }{r^5}(\frac{\kappa q_1 Q^2}{2 r_h}-\frac{r_h^3}{L^2})+\frac{4 Q^4 q_1 \kappa^2}{r^6}-\frac{5 q_1 
	\kappa^2 Q^4}{r^7},
\end{equation}
$C_6$ is determined by applying the condition $f_1(r_h)=0$ as follows,
\begin{equation}\label{C6}
C_6=\frac{24 \kappa  Q^2}{L^2 r_h^2}-\frac{2 \kappa  Q^2}{L^2 r_h}+\frac{7 \kappa   Q^2}{r_h^4}(\frac{\kappa q_1 Q^2}{2 r_h}-\frac{r_h^3}{L^2})-\frac{4 \kappa ^2 Q^4
	q_1 }{r_h^5}+\frac{5 \kappa ^2 Q^4 q_1 }{r_h^6},
\end{equation}
by inserting Eq.(\ref{C6}) in Eq.(\ref{f1}) we have,
\begin{eqnarray}
&&f_1(r)=-\frac{5 \kappa ^2 Q^4 q_1}{r^7} (1-\frac{r^6}{r_h^6})-\frac{\kappa ^2 Q^4 q_1}{2 r^6 r_h^5}(7 r r_h^4+r^5-8 r_h^5)\nonumber \\&&+\frac{\kappa  Q^2 }{L^2 r^5 r_h}\left(2 r^3 r_h-9 r^4+7 r_h^4\right)+\frac{24 \kappa  Q^2 \left(r^2-r_h^2\right)}{L^2 r^3 r_h^2}.
\end{eqnarray}
\section{Holographic Aspects of this Solution}
\label{sec3}
\indent We want to calculate the non-abelian DC conductivity and shear viscosity to entropy density as two important transport coefficients using fluid-gravity to describe the dual of our model.\\
We use the Green-Kubo formula \cite{Policastro2002} for calculating the non-abelian color DC conductivity,    
\begin{equation} \label{kubo2}
\sigma^{ij} (k_{\mu})=-\mathop{\lim }\limits_{\omega \to 0} \frac{1}{\omega } \Im G^{ij}(k_{\mu}).
\end{equation}
The retarded Green's function is calculated using AdS/CFT duality. First, we perturb the gauge field as $A_{\mu} \to A_{\mu}+\tilde{A}_{\mu}$ and then put it into the action. We expand the resulting action \ref{action} up to the second order of the perturbed part. Finally, the Green's function is calculated by taking the second derivative with respect to the value of the gauge field on the boundary \cite{Son}-\cite{Policastro2002},
\begin{equation}
\sigma^{\mu \nu}(\omega)=\frac{1}{i \omega}<J^{\mu}(\omega)J^{\nu}(-\omega)>=\frac{\delta^2 S}{\delta \tilde{A}^0_{\mu} \delta \tilde{A}^0_{\nu}},
\end{equation}
where $\tilde{A}^0_{\nu}$ represents the gauge field perturbation value at the boundary. Boundary has an $SO(2)$ symmetry. Therefore, this condition ensures that conductivity is a scalar quantity.
\begin{equation}
\sigma^{ij}_{ab}=\sigma_{ab}\delta^{ij}
\end{equation}
We consider  the  perturbed part of  gauge  field as $\tilde{A}_x=\tilde{A}_x(r)e^{-i\omega t}$ where $\omega$ should be small - we are on the hydrodynamics regime.\\
By inserting the perturbed part into the action Eq.(\ref{action}) and keeping terms up to second order of $\tilde{A}$ we have,\\
\begin{align}\label{action-2}
S^{(2)}&=-\int d^4x \frac{2\mathcal{Y}}{ f r^2} \Bigg[-f^2 \left((\partial_r\tilde{A}_x^{(1)})^2+(\partial_r\tilde{A}_x^{(2)})^2+(\partial_r\tilde{A}_x^{(3)})^2\right) \nonumber\\
&+\Big((\tilde{A}_x^{(1)})^2+(\tilde{A}_x^{(2)})^2\Big) \left( \omega ^2+h^2\right)+\omega ^2 (\tilde{A}_x^{(3)})^2\Bigg],
\end{align}
in which,
\begin{equation}
\mathcal{Y}=\bigg[ r \left(f' \left(8 q_2-6 q_2 r H'\right)-r \left(q_1-2 q_2 f''\right)\right)+4 q_2 f \left(r^2 H'^2-r^2 H''-2 r H'+1\right)
\bigg].
\end{equation}
By variation of action $S^{(2)}$ with respect to $\tilde{A}_x ^{(1)}$ we have,
 \begin{align}\label{PerA1}
 f \left(f\tilde{A}_x^{(1)'}\right)'+2 \tilde{A}_x^{(1)}\left( h^2+ \omega ^2\right)+\frac{q_2}{q_1 r^2} \Big[ E_0-\frac{4 f}{r}{ (E_1+E_2+E_3)}\Big] =0,     
\end{align}
where,
\begin{align} 
E_0=-4 \tilde{A}_x^{(1)}\left(h^2+\omega ^2\right) \left(r \left(f' \left(4-3 r H'\right)+r f''\right)+2 f \left(\left(r H'-1\right)^2-r^2 H''\right)\right)
\end{align}
\begin{equation}
E_1=r^2 \tilde{A}_x^{(1)'} f' \left(f' \left(4-3 r H'\right)+r f''\right),    
\end{equation}
\begin{align} 
&E_2=r^2 \tilde{A}_x^{(1)''} f \left(f' \left(4-3 r H'\right)+r f''\right) \nonumber\\&+r^2 f \tilde{A}_x^{(1)'} \left(f'' \left(4-3 r H'\right)+f' \left(4 H' \left(r H'-2\right)-7 r H''\right)+r f^{(3)}(r)\right),
\end{align}
\begin{align}
&E_3= 2 f^2 \left(r \tilde{A}_x^{(1)''} \left(\left(r H'-1\right)^2-r^2 H''\right)\right)\nonumber\\&
2\tilde{A}_x^{(1)'}f^2\left(2 \left(r H'-1\right) \left(r^2 H''+1\right)-r^3 H^{(3)}(r)\right).  
\end{align}
The $\tilde{A}_x^{(2)}$ part is the same as  $\tilde{A}_x^{(1)}$.\\
The $\tilde{A}_x^{(3)}$ part is as follows,  
 \begin{align}\label{PerA3}
 &q_2 \tilde{A}_x^{(3)} r  \omega ^2 \bigg( 2 r f' \left(3 r H'-4\right)-2r^2 f''-4 f\left(1-2 r H'+r^2 H'^2\right)-r^2
 H''\bigg)\nonumber\\&+q_2 f r^2 \Bigg(\tilde{A}_x^{(3)''}  \Big(2f'\left(-4 + 3 r H'\right)-2rf''\Big)+r^3 q_1 \Bigg( f \left(f
 \tilde{A}_x^{(3)'}\right)'+\omega ^2\tilde{A}_x^{(3)}\Bigg)  \nonumber\\&-2\tilde{A}_x^{(3)'} r^2 f \left(f'' \left(4-3 r H'\right)+f' \left(4 H' \left(r H'-2\right)-7 r H''\right)+r f^{(3)}
 \right)\nonumber\\&+4  f^2 \left( r \tilde{A}_x^{(3)''}\left(2r H'-1-r^2 H'^2+r^2 H''\right)+\tilde{A}_x^{(3)'} \left(2 r^2 H''+2-2 H'(r+r^3 H'')+r^3
 H^{(3)}(r)\right) \right)\Bigg)=0    
 \end{align}
By utilizing the following relationships: $f_0\sim4\pi f_0'(r_h)(r-r_h)$ and $f_1\sim4\pi f_1'(r_h)(r-r_h)$, we can determine the solution for Eq.(\ref{PerA1}) and Eq.(\ref{PerA3}) near the event horizon. Since $\tilde{A}_x^{(a)}$ should vanish on the event horizon, we consider $\tilde{A}_x^{(a)}$ as follows,
\begin{align}
\tilde{A}_x^{(a)}\sim (r-r_h)^{z_a} \, , \qquad a=1,2,3
\end{align}
where,
\begin{align}\label{z12}
z_1&=z_2=\pm i \frac{\sqrt{h(r_h)^2+\omega ^2}}{4 \pi T}  \\
\label{z3}
z_3&=\pm i \frac{\omega }{4 \pi T}.
\end{align}
The Hawking temperature of the black brane $T$ is as follows,
\begin{equation}
T=\frac{1}{2 \pi} \Big[ \frac{1}{\sqrt{g_{rr}}}\frac{d}{dr}\sqrt{-g_{tt}}\Big]\Bigg|_{r=r_h}=\frac{e^{-H(r_h)} f'(r_h)}{4 \pi}.
\end{equation} 
For solving the $\tilde{A}_x^{(a)}$  from event horizon to boundary, we consider this ansatz as following, 
\begin{align}\label{EOMA1}
\tilde{A}_x^{(1)}=\tilde{A}^{(1)}_{\infty}\Big(\frac{-3f}{\Lambda r^2}\Big)^{z_1}\Big(1+i\omega b_1(r)+\cdots\Big) ,
\end{align}
\begin{align}\label{EOMA2}
\tilde{A}_x^{(2)}=\tilde{A}^{(2)}_{\infty}\Big(\frac{-3f}{\Lambda r^2}\Big)^{z_2}\Big(1+i\omega b_2(r)+\cdots\Big) ,
\end{align}
\begin{align}\label{EOMA3}
\tilde{A}_x^{(3)}=\tilde{A}^{(3)}_{\infty}\Big(\frac{-3f}{\Lambda r^2}\Big)^{z_3}\Big(1+i\omega b_3(r)+\cdots\Big) ,
\end{align}
in which $\tilde{A}^{(a)}_{\infty}$ represents the value of fields at the boundary, and $z_i$'s correspond to the negative signs in Eq.(\ref{z12}) and Eq.(\ref{z3}). We select the ingoing mode of near horizon because the outgoing mode does not exist from the event horizon of black hole.\\
By applying the regularity condition of $\tilde{A}_x^{(3)}$ on the event horizon, the term $b_3(r)$ in Eq.(\ref{EOMA3}) can be determined as follows,
\begin{equation}
b_3(r)= \int^r\left(\frac{2}{u}-\frac{f'}{f}+\frac{C_7 u^2}{fN}\right)du+C_8,    
\end{equation}
where,
\begin{align}
N=2 u q_2 f' \left(4-3 u H'\right)-u^2 \left(q_1-2 q_2 f''\right)+4 q_2 f \left(u H'-1\right)^2-4 q_2 f u^2 H''
\end{align}
The solution of $b_3(r)$ to first order of $q_2$ is as following, 
\begin{align}\label{b3}
&b_3(r)=\int^r\left(\frac{2}{u}-\frac{f_0'}{f_0}-\frac{C_7}{q_1  f_0}\right)du\nonumber\\&+ q_2 \int^r\  \Bigg(\frac{f_1 f_0'}{f_0^2}-\frac{8 C_1 f_0'}{q_1^2 u f_0}-\frac{2 C_1
 f_0'''}{q_1^2 f_0}-\frac{f_1'}{f_0'}+\frac{C_1 f_1}{q_1 f_0^2}-\frac{4
	C_1}{q_1^2 u^2}    \Bigg) du,
\end{align}
where $C_7$ and $C_8$ are integration constants.\\ 
We find the solution of $b_3(r)$ near the event horizon as,\\
\begin{equation}\label{C4}
b_3 \approx  \Bigg(-1-\frac{C_7}{q_1 f_0'(r_h)}+q_2\bigg( \frac{ f_1'(r_h)}{f_0'(r_h)}-\frac{8 C_7 }{q_1}-\frac{2 C_7 f_0'''(r_h)}{q_1 f_0'(r_h)}+\frac{C_7 f_1'(r_h)}{q_1 f_0'^2(r_h)}\bigg)\Bigg) \log(r-r_h)+\text{finite terms}.
\end{equation}
The solution should be regular on the event horizon. Therefore, $C_7$ is determined by demanding the following condition,
\begin{equation}
C_7=\frac{1-q_2 \frac{f_1'(r_h)}{f_0'(r_h)}}{\frac{-1}{q_1 f_0'(r_h)}-\frac{8q_2}{q_1}-\frac{2 q_2 f_0'''(r_h)}{q_1 f_0'(r_h)}+\frac{ q_2 f_1'(r_h)}{q_1 f_0'^2(r_h)}},
\end{equation}
$C_7$ is found up to first order of $q_2$ it can be expressed as,
\begin{equation}
C_7=-q_1 f_0'(r_h)+q_1 q_2 \left(f_0'(r_h)^2
\left(8-\frac{f_1'(r_h)^2}{f_0'^2(r_h)}\right)+2 f_0'(r_h)
f_0''(r_h)+f_1'(r_h)\right).
\end{equation}
Considering the solution of $\tilde{A}_x^{(3)}$ in Eq.(\ref{action-2}) and variation with respect to $\tilde{A}^{(3)}_{\infty}$, Green's function  can be read as,
\begin{align} \label{Green1}
& G_{xx}^{(33)} (\omega ,\vec{0})=-i \omega \frac{C_7}{q_1 f_0'(r_h)}=-i \omega+i \omega q_2\left(f_0'(r_h)
\left(8-\frac{f_1'(r_h)^2}{f_0'^2(r_h)}\right)+2 
f_0''(r_h)+\frac{f_1'(r_h)}{f_0'(r_h)}\right).
\end{align}
The conductivity is as following, 
\begin{eqnarray}\label{sigma33}
\sigma_{xx}^{(33)}=-\mathop{\lim }\limits_{\omega \to 0} \frac{1}{\omega } \Im G^{ij}(k_{\mu}) =1- q_2\left(f_0'(r_h)
\left(8-\frac{f_1'(r_h)}{f_0'^2(r_h)}\right)+2 
f_0''(r_h)+\frac{f_1'(r_h)}{f_0'(r_h)}\right)
\end{eqnarray}
by substituting $f_0(r)$ , $f_1(r)$  in above equation we have, 
\begin{eqnarray}
\sigma_{xx}^{(33)} =1-4 q_2\Bigg(\frac{6  r_h}{L^2}-\frac{ \kappa  Q^2 q_1  (r_h-1)}{r_h^4}\Bigg).
\end{eqnarray}
It shows that the conductivity bound is violated for the non-abelian non-minimal $R F^2$ black brane theory for $\Big(\frac{6  r_h}{L^2}-\frac{ \kappa  Q^2 q_1  (r_h-1)}{r_h^4}\Big)>0$. In the limit of $q_2 \to 0$,  we have,\\
\begin{eqnarray}
\sigma_{xx}^{(33)} =1.
\end{eqnarray}
For calculating the ratio of shear viscosity to entropy density, we follow the method outlined in [\cite{Hartnoll:2016tri}].\\
We must perturb the metric as $g_{\mu \nu} \to g_{\mu \nu} +\delta g_{xy}$. Where $\delta g_{xy} = \frac{r^2}{L^2} \phi(r )e^{i\omega t}$ and we consider $\omega=0$.  $\frac{\eta}{s}$  is read as follows,\\
\begin{equation}
\frac{\eta}{s}=\frac{1}{4 \pi} \phi(r_h)^2.
\end{equation}
We insert the perturbed part of the metric into the action and expand the action up to the second order of $\phi$. By taking the variation of the resulting action with respect to $\phi$ we have,
\begin{eqnarray}
r f' \phi '-r f H' \phi'+3 f \phi'+r f \phi''=0
\end{eqnarray}
The solution of the $\phi(r)$ is as follows,
\begin{align}\label{phi}
\phi (r)=C_5 + C_6\int^{r}\frac{e^{2 H(u)} }{u^3 f(u)}du.
\end{align}
By inserting Eq. (\ref{f}) - Eq.(\ref{H}) in Eq. (\ref{phi}), we find the solution of $\phi$ in the leading order of $q_2$ as follows,
\begin{align}
\phi (r)=\phi_0 (r)+q_2 \phi_1 (r)=C_5+ C_6\int ^{r}\frac{du}{u^3 f_0(u) }+C_6 q_2  \int^r\frac{ f_0 H_1-f_1}{u^3 f_0^2}du.
\end{align}
We calculate the solution of $\phi_0 (r)$ near the event horizon, we obtain,
\begin{eqnarray}
&&\phi(r)=\phi_0 (r)+q_2\phi_1 (r)= C_5+  \frac{C_6}{4 \pi T f_0'(r_h) r_h^3}\log(r-r_h)\nonumber\\&&+C_6 q_2 \Bigg(\frac{2 H_1(r_h)  }{r_h^3 f_0'(r_h)}-\frac{f_1'(r_h)  }{r_h^3 f_0'(r_h)^2}\Bigg)\log(r-r_h),
\end{eqnarray}
$\phi_0 (r)$ should be regular on the event horizon so $C_6=0$ and we set $C_5=1$ for normalization of $\phi_0 (r)$.
\begin{align}
\phi_0 (r)=1
\end{align}
$\phi (r)$ up to first order of $q_2$ is as,
\begin{align}
\phi_1 (r)=0,
\end{align}
then, the solution of $\phi(r)$ is as follows,
\begin{align}
\phi(r)=\phi_0(r)+q_2 \phi_1(r)=1.
\end{align}
Finally, the ratio of shear viscosity to entropy density up to first order of $q_2$ is as,
\begin{equation}
\frac{\eta}{s}=\frac{1}{4 \pi}.
\end{equation}
Kovtun, Son and Starinets (KSS) bound states that $\frac{\eta}{s} \geq \frac{1}{4 \pi} $ for all  quantum field theories. This bound is saturated for Einstein-Hilbert gravity with field theory dual and it is violated for higher derivative gravities. The KSS bound is preserved for this model.
 \section{Conclusion}

\noindent In this paper, we introduced non-minimal $RF^{(a)}_{\mu \alpha }F^{(a)\mu \alpha} $ black brane solution in AdS spacetime in four dimensions. Since this model does not have an analytical solution, we solve it perturbatively in terms of the non-minimal coupling $q_2$. We investigate the duality of this model via fluid-gravity duality by calculating the color non-abelian DC conductivity and the ratio of shear viscosity to entropy density, which are two important transport coefficients. Our results show that the conductivity bound is violated here for some values of $q_2$. The reason is the effect of the $RF^{(a)}_{\mu \alpha }F^{(a)\mu \alpha} $ term and in the limit of $q_2 \to 0$ the conductivity bound is saturated and it recovers the result of the Yang-Mills model. The term $RF^{(a)}_{\mu \alpha }F^{(a)\mu \alpha} $ also does not affect the value of $\frac{\eta}{s}$ up to the first order of $q_2$. Since $\frac{\eta}{s}$ is inversely proportional to the square of the coupling on the field theory side, it implies that the dual of our model is equivalent to the dual of Einstein-Hilbert gravity, as the field theory coupling is identical.

\vspace{1cm}
\noindent {\large {\bf Acknowledgment} } Author would like to thank Shahrokh Parvizi, Komeil Babaei and Faramarz Rahmani for useful comments and suggestions. We also gratefully thank to the MPLA Referee for the constructive comments and recommendations which definitely help to improve the readability and quality of the manuscript.\\ 

\noindent {\large {\bf Data availability statement} }\\

All data that support the findings of this study are included within the article.



\begin{thebibliography}{}

\bibitem{Hawking1983}
Hawking, S.W., Page, D.N. 
 Commun.Math. Phys. 87, 577–588 (1983). https://doi.org/10.1007/BF01208266.

 \bibitem{Maldacena}
J. M. Maldacena, 
 Int.\ J.\ Theor.\ Phys.\  {\bf 38} (1999) 1113 [Adv.\ Theor.\ Math.\ Phys.\  {\bf 2} (1998) 231] [hep-th/9711200].


\bibitem{Gubser:1998bc}
S.~S.~Gubser, I.~R.~Klebanov and A.~M.~Polyakov,
Phys. Lett. B \textbf{428}, 105-114 (1998)
doi:10.1016/S0370-2693(98)00377-3
[arXiv:hep-th/9802109 [hep-th]].


\bibitem{Witten:1998qj}
E.~Witten,
Adv. Theor. Math. Phys. \textbf{2}, 253-291 (1998)
doi:10.4310/ATMP.1998.v2.n2.a2
[arXiv:hep-th/9802150 [hep-th]].




  \bibitem{Aharony}
  O. Aharony, S.~S.~Gubser, J.~M.~Maldacena, H.~Ooguri and Y.~Oz,
  Phys.\ Rept.\  {\bf 323}, 183 (2000)
  [hep-th/9905111].


  

  \bibitem{Son}
  D.T. Son, 
  Nuclear Physics B (Proc. Suppl.) 192--193 (2009) 113--118.
  

 
 
 \bibitem{Son2007}
  D.~T.~Son and A.~O.~Starinets,
 Ann.\ Rev.\ Nucl.\ Part.\ Sci.\  {\bf 57}, 95 (2007)
  [arXiv:0704.0240 [hep-th]].



\bibitem{Policastro2002}
G.~Policastro, D.~T.~Son and A.~O.~Starinets,
JHEP {\bf 0209}, 043 (2002)
[hep-th/0205052].





  
  \bibitem{Policastro2001}
  G.~Policastro, D.~T.~Son and A.~O.~Starinets,
  Phys.\ Rev.\ Lett.\  {\bf 87}, 081601 (2001)
  [hep-th/0104066].
 


  
  \bibitem{Kovtun2012}
  P. Kovtun,
  J.\ Phys.\ A {\bf 45} (2012) 473001[arXiv:1205.5040 [hep-th]].
  
  \bibitem{Bhattacharyya}
  S.~Bhattacharyya, V.~E.~Hubeny, S.~Minwalla and M.~Rangamani,
  JHEP {\bf 0802}, 045 (2008)
  [arXiv:0712.2456 [hep-th]].
  
  
  
  
  
  
  
\bibitem{Balakin:2015gpq}
A.~B.~Balakin, J.~P.~S.~Lemos and A.~E.~Zayats,
Phys. Rev. D \textbf{93}, no.2, 024008 (2016)
doi:10.1103/PhysRevD.93.024008
[arXiv:1512.02653 [gr-qc]].







\bibitem{Edery:2018jyp}
A.~Edery and Y.~Nakayama,
Phys. Rev. D \textbf{98}, no.6, 064011 (2018)
doi:10.1103/PhysRevD.98.064011
[arXiv:1807.07004 [hep-th]].
 
 
 
 \bibitem{Bamba:2008ja}
 K.~Bamba and S.~D.~Odintsov,
 JCAP \textbf{04}, 024 (2008)
 doi:10.1088/1475-7516/2008/04/024
 [arXiv:0801.0954 [astro-ph]].
 
 
 \bibitem{Horndeski:1978ca}
 G.~W.~Horndeski,
 Phys. Rev. D \textbf{17}, 391-395 (1978)
 doi:10.1103/PhysRevD.17.391
 
 
 
 \bibitem{Mueller-Hoissen:1988cpx}
 F.~Mueller-Hoissen and R.~Sippel,
 Class. Quant. Grav. \textbf{5}, 1473 (1988)
 
 \bibitem{Balakin:2007xq}
 A.~B.~Balakin, S.~V.~Sushkov and A.~E.~Zayats,
 Phys. Rev. D \textbf{75}, 084042 (2007)
 doi:10.1103/PhysRevD.75.084042
 [arXiv:0704.1224 [gr-qc]].
 
 
 \bibitem{Balakin:2010ar}
 A.~B.~Balakin, J.~P.~S.~Lemos and A.~E.~Zayats,
 Phys. Rev. D \textbf{81}, 084015 (2010)
 
 \bibitem{Balakin:2007am}
 A.~B.~Balakin, V.~V.~Bochkarev and J.~P.~S.~Lemos,
 Phys. Rev. D \textbf{77}, 084013 (2008)
 doi:10.1103/PhysRevD.77.084013
 [arXiv:0712.4066 [gr-qc]].
 
 
 \bibitem{Balakin:2015oea}
 A.~B.~Balakin and A.~E.~Zayats,
 Int. J. Mod. Phys. D \textbf{24}, no.09, 1542009 (2015)
 doi:10.1142/S0218271815420092
 [arXiv:1506.05236 [gr-qc]].
 
 \bibitem{Balakin:2006gv}
 A.~B.~Balakin and A.~E.~Zayats,
 Phys. Lett. B \textbf{644}, 294-298 (2007)
 doi:10.1016/j.physletb.2006.12.002
 [arXiv:gr-qc/0612019 [gr-qc]].
 
 
 
 \bibitem{Brigante:2007nu}
 M.~Brigante, H.~Liu, R.~C.~Myers, S.~Shenker and S.~Yaida,
 Phys. Rev. D \textbf{77}, 126006 (2008)
 doi:10.1103/PhysRevD.77.126006
 [arXiv:0712.0805 [hep-th]].
 
  
\bibitem{Sadeghi:2015vaa}
M.~Sadeghi and S.~Parvizi,
Class. Quant. Grav. \textbf{33}, no.3, 035005 (2016)
doi:10.1088/0264-9381/33/3/035005
[arXiv:1507.07183 [hep-th]].


\bibitem{Parvizi:2017boc}
S.~Parvizi and M.~Sadeghi,
Eur. Phys. J. C \textbf{79}, no.2, 113 (2019)
doi:10.1140/epjc/s10052-019-6631-9
[arXiv:1704.00441 [hep-th]].


\bibitem{Bravo-Gaete:2020lzs}
M.~Bravo-Gaete and F.~F.~Santos,
Eur. Phys. J. C \textbf{82}, no.2, 101 (2022)
doi:10.1140/epjc/s10052-022-10064-y
[arXiv:2010.10942 [hep-th]].



\bibitem{Bravo-Gaete:2021hlc}
M.~Bravo-Gaete and M.~M.~Stetsko,
Phys. Rev. D \textbf{105}, no.2, 024038 (2022)
doi:10.1103/PhysRevD.105.024038
[arXiv:2111.10925 [hep-th]].  

\bibitem{Bravo-Gaete:2022lno}
M.~Bravo-Gaete, F.~F.~Santos and H.~Boschi-Filho,
Phys. Rev. D \textbf{106}, no.6, 066010 (2022)
doi:10.1103/PhysRevD.106.066010
[arXiv:2201.07961 [hep-th]]. 




\bibitem{Grozdanov:2015qia}
S.~Grozdanov, A.~Lucas, S.~Sachdev and K.~Schalm,
Phys. Rev. Lett. \textbf{115}, no.22, 221601 (2015)
doi:10.1103/PhysRevLett.115.221601
[arXiv:1507.00003 [hep-th]]. 



  \bibitem{Donos:2014cya}
  A.~Donos and J.~P.~Gauntlett,
  JHEP {\bf 1411}, 081 (2014)
  doi:10.1007/JHEP11(2014)081
  [arXiv:1406.4742 [hep-th]].


\bibitem{Sadeghi:2021qou}
M.~Sadeghi,
doi:10.1007/s12648-022-02317-z
[arXiv:2111.12916 [hep-th]].


\bibitem{Sadeghi:2022mog}
M.~Sadeghi,
[arXiv:2203.05023 [hep-th]].


\bibitem{Sert:2020vmq}
\"O.~Sert and F.~\c{C}elikta\c{s},
Eur. Phys. J. C \textbf{80}, no.7, 653 (2020)
doi:10.1140/epjc/s10052-020-8237-7
[arXiv:2004.06769 [gr-qc]].


\bibitem{Lambiase:2008zz}
G.~Lambiase, S.~Mohanty and G.~Scarpetta,
JCAP \textbf{07}, 019 (2008)
doi:10.1088/1475-7516/2008/07/019




\bibitem{Dereli:2011mk}
T.~Dereli and O.~Sert,
Eur. Phys. J. C \textbf{71}, 1589 (2011)
doi:10.1140/epjc/s10052-011-1589-2
[arXiv:1102.3863 [gr-qc]].






\bibitem{Balakin:2005fu}
A.~B.~Balakin and J.~P.~S.~Lemos,
Class. Quant. Grav. \textbf{22}, 1867-1880 (2005)
doi:10.1088/0264-9381/22/9/024
[arXiv:gr-qc/0503076 [gr-qc]].


\bibitem{Shepherd:2015dse} 
 B.~L.~Shepherd and E.~Winstanley,
  Phys.\ Rev.\ D {\bf 93}, no. 6, 064064 (2016)
  doi:10.1103/PhysRevD.93.064064
  [arXiv:1512.03010 [gr-qc]].



\bibitem{Myers:2009ij}
R.~C.~Myers, M.~F.~Paulos and A.~Sinha,
JHEP \textbf{06}, 006 (2009)
doi:10.1088/1126-6708/2009/06/006
[arXiv:0903.2834 [hep-th]].




\bibitem{Dey:2015poa}
A.~Dey, S.~Mahapatra and T.~Sarkar,
JHEP \textbf{01}, 088 (2016)
doi:10.1007/JHEP01(2016)088
[arXiv:1510.00232 [hep-th]].

\bibitem{Dey:2015ytd}
A.~Dey, S.~Mahapatra and T.~Sarkar,
Phys. Rev. D \textbf{94}, no.2, 026006 (2016)
doi:10.1103/PhysRevD.94.026006
[arXiv:1512.07117 [hep-th]].


\bibitem{Mahapatra:2016dae}
S.~Mahapatra,
JHEP \textbf{04}, 142 (2016)
doi:10.1007/JHEP04(2016)142
[arXiv:1602.03007 [hep-th]].


  
  \bibitem{Hartnoll:2016tri}
  S.~A.~Hartnoll, D.~M.~Ramirez and J.~E.~Santos,
  JHEP \textbf{03}, 170 (2016)
  doi:10.1007/JHEP03(2016)170
  [arXiv:1601.02757 [hep-th]].
  
   
  
  
\end{thebibliography}
\end{document}